\documentclass[prd,reprint,showpacs,showkeys]{revtex4-1}
\usepackage{amsfonts}
\usepackage{amssymb}
\usepackage{amsmath}
\usepackage{graphicx}
\usepackage{float}
\usepackage{subcaption}
\usepackage[font={footnotesize,it}]{caption}

\begin{document}

\title{Regularization of the Reissner-Nordstr\"{o}m black hole}
\author{S. Habib Mazharimousavi}
\email{habib.mazhari@emu.edu.tr}
\author{M. Halilsoy}
\email{mustafa.halilsoy@emu.edu.tr}
\affiliation{Department of Physics, Faculty of arts and sciences, Eastern Mediterranean
	University, Famagusta, North Cyprus, via Mersin 10, Turkey. }

\date{\today}

\begin{abstract}
An inner de Sitter region is glued smoothly and consistently with an outer
Reissner-Nordstr\"{o}m (RN) spacetime on a spherical thin-shell. Mass and
charge of the outer RN spacetime are defined by the de Sitter and shell
parameters. Radius of the shell plays the role of a cut-off which by virtue
of regular de Sitter inside removes the singularity at $r=0.$ The topology
of inner de Sitter with the radius of the thin-shell becomes compact. For
stability the perturbed shell is shown to satisfy a modified polytropic
equation of state which has vanishing mass and pressure on the unperturbed
shell as dictated by the junction conditions.
\end{abstract}

\maketitle

\section{Introduction}

Since the inception of the cut and paste technique following the seminal
work of Israel's junction conditions \cite{Israel} the topic of thin-shells
has been popularized extensively. Application of thin-shells to wormholes 
\cite{MV} in general relativity has been another major topic that found vast
applications. In that construction (preferably) two asymptotically flat
spacetimes are glued at a minimal radius that defines the throat of the
wormhole \cite{MT}. Through that throat an observer passes from one universe
to the other easily. A wormhole may connect two black holes which may be
interpreted in the language of modern physics as entanglement \cite%
{Maldacena}. It should also be reminded that the existence of a minimum
radius lead Einstein and Rosen to interpret a wormhole as a geometrical
model of a particle \cite{ER}. Very special spacetimes satisfy Israel's
junction conditions \cite{Israel} to be glued smoothly \cite{Vilenkin1,
Vilenkin2, Zaslavskii}. In \cite{Vilenkin1, Vilenkin2} inner flat /
Minkowski spacetime was glued to the outer extremal RN. However, Zaslavskii
in \cite{Zaslavskii} has shown that Minkowski spacetime can not be glued
smoothly to extremal RN but instead Bertotti-Robinson spacetime was
successfully glued to extremal RN black hole at its horizon.

In this paper we glue an inner de Sitter with an outer Reissner-Nordstr\"{o}%
m (RN) spacetime on a spherical shell that satisfies the junction conditions
of smooth match The reasons and advantages for such an option have already
been explained in details by Lemos and Zanchin \cite{9} and Uchihata et. al 
\cite{10}. Our analysis is closely related to their works while the
distinction from theirs will be justified below. Our choice, is made such
that as in \cite{9} no energy momentum tensor exists on the interface
hypersurface \cite{Zaslavskii} i.e., the mass = the pressure = 0 at
equilibrium. In this sense our work is different from that of Frolov, et. al 
\cite{Frolov}. (Frolov, in a more recent work, presented also a generic
approach to the non-singular models of black holes in static spherically
symmetric spacetime in four and higher dimensions \cite{Frolov2}.) We employ
the matching conditions by replacing the singular inside of a RN black hole
with a regular de Sitter spacetime with a compact topological structure. In
the literature there are both regular black holes \cite{RBH} as well as
regularization methods \cite{Regularization} which involve a change in
topology of the spacetime. Specifically, the stability analysis of the shell
distinguishes our work in the present study from that of \cite%
{Regularization}. The external RN spacetime which has parameters mass $M$
and charge $Q$ are determined from the satisfaction of the boundary
conditions required for a smooth match. Stated otherwise, the mass and
charge are defined 'from geometry' in accordance with Wheeler's
geometrodynamics \cite{Wheeler}. Regularization is to be understood in the
sense that is reminiscent of some renormalization / regularization
techniques that were used in field theory. The aim in those techniques was
to eliminate divergences in field theory. In doing this, experimental values
of particles, such as charge, mass, magnetic moment etc. \ were used as
guidelines. Insertion of measured quantities into the theory played major
role in choosing the cut-offs. As a result finite quantities emerged from
the divergent ones, as a physical requirement. In this study we shall insert
a thin-shell of radius $R_{0}\neq 0$, as our cut-off to eliminate the
singularity at the origin. The radius $R_{0}$ of the shell must be finely
tuned since it will be related to the mass $M$, charge $Q$ and the
cosmological constant $\ell .$

In general relativity also singularities, i.e., diverging curvature
invariants lie at the heart of gravitational theory. Most black holes admit
singularities at their center which make invariants divergent. The worst of
such singularities is the spacelike ones as encountered in the Schwarzschild
black hole. Addition of electric charge (i.e., the RN solution) makes the
central singularity time-like, which is the subject matter of the present
article. By cutting the central singularity and pasting a regular de Sitter
spacetime we get rid of the $r=0$ singularity. In turn, the shell must
satisfy certain conditions, especially upon perturbation for stability
requirement a fluid energy-momentum arises naturally. This is in the form of
a modified polytropic fluid whose energy density and transverse pressures
satisfy the conservation law. We discuss briefly the physical properties of
such a fluid. Being highly nonlinear we choose a particular case and confine
the argument to the vicinity of the static shell. Before perturbation the
shell may be taken as a false vacuum state as in the field theory in which
the surface energy-momentum of the fluid vanishes i.e., energy density $%
\sigma $ and pressure $p$ are both zero. The total energy analysis after the
perturbation suggests an energy zone that makes the shell and therefore our
model, stable against linear radial perturbations.

Organization of the paper is as follows. In Section II we introduce our
model of gluing inner de Sitter with the outer RN metrics. Energy-momentum
and Maxwell equations on the shell are discussed in Section III. Section IV
analyses the stability of the model. Our Conclusion and Discussion appears
in Section V.

\section{The model}

In $3+1-$dimension, let's consider the following static, spherically
symmetric spacetimes 
\begin{equation}
ds^{2}=-f_{i}\left( r_{i}\right) dt_{i}^{2}+\frac{dr_{i}^{2}}{f_{i}\left(
r_{i}\right) }+r_{i}^{2}\left( d\theta _{i}^{2}+\sin ^{2}\theta _{i}d\phi
_{i}^{2}\right)
\end{equation}%
for inside ($i=1$) and outside ($i=2$) of a timelike shell defined by $%
F:=r-R_{0}=0$ where $R_{0}$ is the constant radius of the shell. Following
the Israel junction formalism \cite{Israel}, the induced metric on the shell
is found to be 
\begin{equation}
ds^{2}=-d\tau ^{2}+R_{0}^{2}\left( d\theta ^{2}+\sin ^{2}\theta d\phi
^{2}\right) .
\end{equation}%
The energy momentum tensor components on the shell are from $S_{\mu }^{\nu
}=diag\left( \sigma _{0},p_{0},p_{0}\right) $ such that 
\begin{equation}
\sigma _{0}=-\frac{1}{4\pi G}\left( \frac{\sqrt{f_{2}\left( R_{0}\right) }-%
\sqrt{f_{1}\left( R_{0}\right) }}{R_{0}}\right)
\end{equation}%
and%
\begin{multline}
p_{0}=\frac{1}{8\pi G}\left( \frac{f_{2}^{\prime }\left( R_{0}\right) }{2%
\sqrt{f_{2}\left( R_{0}\right) }}-\frac{f_{1}^{\prime }\left( R_{0}\right) }{%
2\sqrt{f_{1}\left( R_{0}\right) }}\right. + \\
\left. \frac{\sqrt{f_{2}\left( R_{0}\right) }-\sqrt{f_{1}\left( R_{0}\right) 
}}{R_{0}}\right)
\end{multline}%
where a prime means $\frac{d}{dr}$ at $r=R_{0}.$ Next, we set 
\begin{equation}
f_{1}=1-\frac{r_{1}^{2}}{\ell ^{2}}
\end{equation}%
and%
\begin{equation}
f_{2}=1-\frac{2M}{r_{2}}+\frac{Q^{2}}{r_{2}^{2}}
\end{equation}%
as representatives of the inner ($f_{1}$) and the outer ($f_{2}$)
spacetimes, respectively \cite{9}. Our aim is to glue the two spacetimes
smoothly such that $\sigma _{0}$ and $p_{0}$ are determined on the shell:
interestingly both vanish. For this we impose $f_{1}\left( R_{0}\right)
=f_{2}\left( R_{0}\right) $ and $f_{1}^{\prime }\left( R_{0}\right)
=f_{2}^{\prime }\left( R_{0}\right) $ which leads to%
\begin{equation}
M=\frac{2R_{0}^{3}}{\ell ^{2}}
\end{equation}%
and%
\begin{equation}
Q^{2}=\frac{3R_{0}^{4}}{\ell ^{2}}.
\end{equation}%
Thus, geometrical conditions of continuity of the metric and its first
derivative automatically determine these fine-tuning conditions that play
crucial role in the problem. Let us add that in this identification the
dimensions of $M$ and $Q$ are same as $R_{0}$ and $\ell .$ For a double
horizon case we must have the condition $R_{0}>\frac{\sqrt{3}}{2}\ell $
satisfied. The choice $R_{0}=\frac{\sqrt{3}}{2}\ell $ will obviously
correspond to the extremal RN and $R_{0}<\frac{\sqrt{3}}{2}\ell $ will give
rise to no horizon case. It is observed that for a nontrivial matching the
limit $R_{0}\rightarrow 0,$ must be excluded. In Fig. 1 we plot $f\left(
r\right) =f_{1}\left( r\right) \Theta \left( R_{0}-r\right) +f_{2}\left(
r\right) \Theta \left( r-R_{0}\right) $ in which $\Theta \left( .\right) $
stands for the Heaviside step function, for different values of $Q$ and $M$
(and consequently $R_{0}$ and $\ell ^{2}$).

\begin{figure}[h]
\includegraphics[width=70mm,scale=0.7]{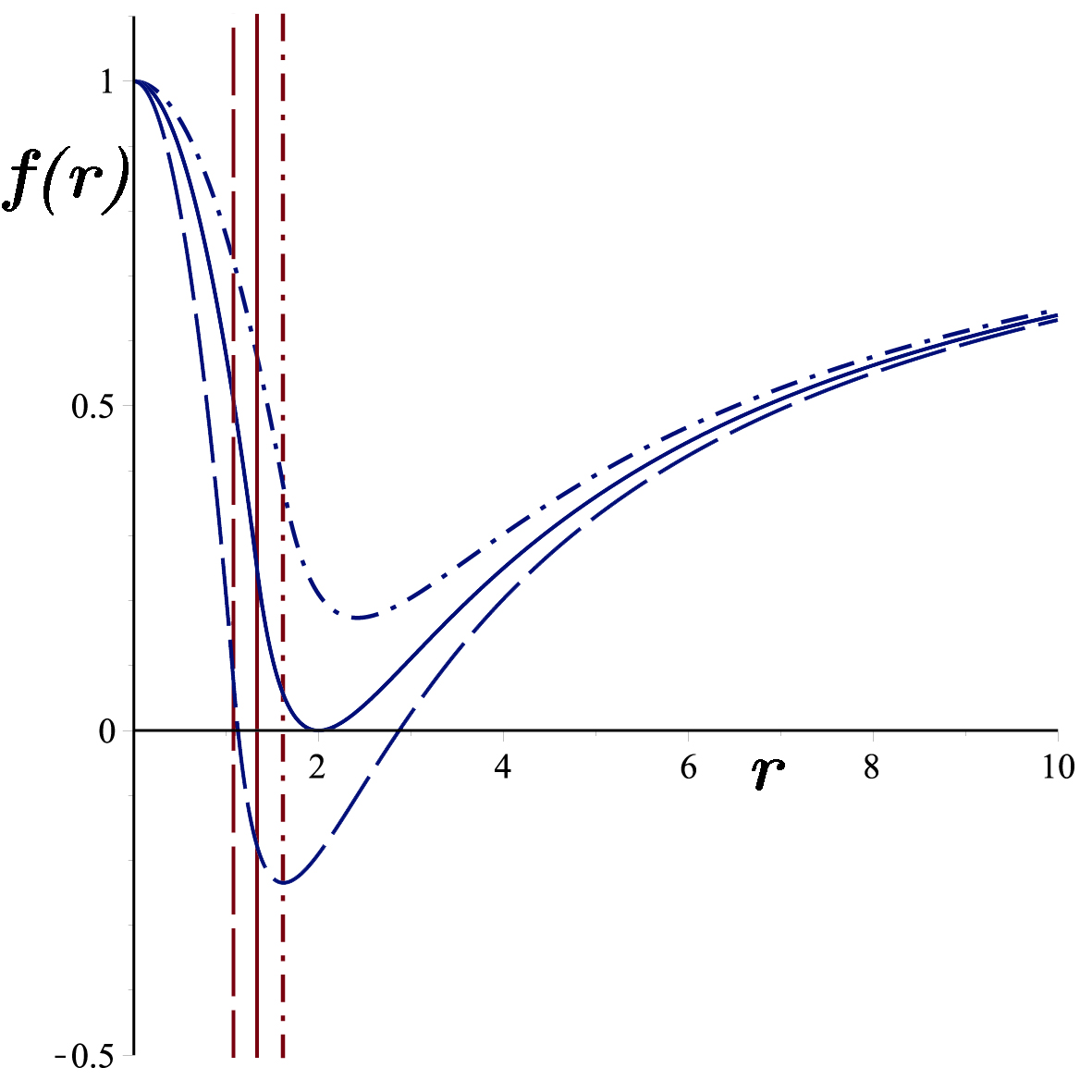}
\caption{{}The metric function $f\left( r\right) =f_{1}\left( r\right)
\Theta \left( R_{0}-r\right) +f_{2}\left( r\right) \Theta \left(
r-R_{0}\right) $ versus $r$ for different values of $Q$ and $M=2.$ From top
to bottom: $Q=2.2,2$ and $1.8$ (or $\left( R_{0}=1.61,\ell ^{2}=4.20\right)
,\left( R_{0}=4/3,\ell ^{2}=64/27\right) $ and $\left( R_{0}=1.08,\ell
^{2}=1.26\right) $. The vertical lines are the locations of the interface
shell, i.e. $r=R_{0}$.}
\end{figure}

\begin{figure}[h]
\includegraphics[width=70mm,scale=0.7]{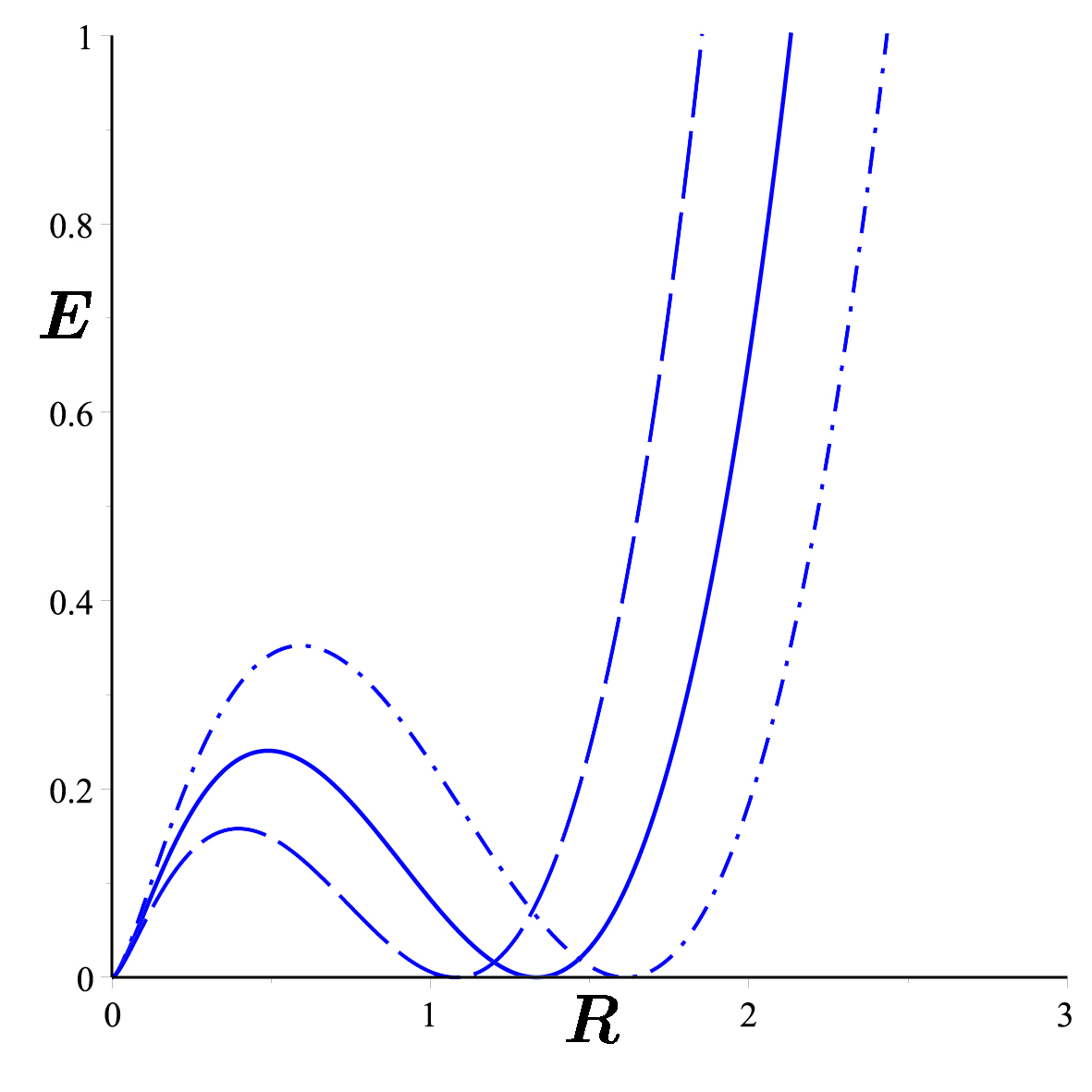}
\caption{{}$E$ vs $R$ for $\protect\omega =1$, $\protect\nu =\frac{1}{2}$
and $R_{0}$ corresponding to Fig. 1. The energy density at $R=R_{0}$ is zero
and so is $p.$ Note that Dash-dot, Solid and Long-dash curves in Fig. 1 and
2 are corresponding to each other.}
\end{figure}

Our Fig. 1 may be compared with the Fig. 4 of Ref. \cite{Frolov} to see the
difference with the de-Sitter-Schwarzschild matching. The thin-shell, or
transition layer in the terminology of Frolov et. al. \cite{Frolov} has a
total non-zero mass / energy whereas in our case by virtue of (7) and (8)
and the definition of the surface energy-momentum tensor $S_{\mu }^{\nu
}=diag\left( \sigma _{0},p_{0},p_{0}\right) $ we have $S_{\mu \nu }=0$. Upon
perturbation, as we shall show below we shall have $S_{\mu \nu }\neq 0$.
Since $r<R_{0}$ for inside and $R_{0}$ can be chosen arbitrary in terms of $%
\ell ,$ say $R_{0}=\alpha \ell ,$ where $\alpha \geq \frac{\sqrt{3}}{2},$
the topology of de Sitter is adjusted accordingly.

From $1-\frac{2M}{r}+\frac{Q^{2}}{r^{2}}=0$ we have $r_{\pm }=M\pm \sqrt{%
M^{2}-Q^{2}}.$ With the substitutions (7), (8) and $R_{0}=\alpha \ell $ we
obtain $r_{\pm }=R_{0}\left( 2\alpha ^{2}\pm \alpha \sqrt{4\alpha ^{2}-3}%
\right) .$ Let us discuss the following cases:

i) For $\alpha ^{2}=3/4$ we have $r_{+}=r_{-}=\frac{3}{2}R_{0}$ which
implies that the horizon lies outside the shell. Upon substitution of $\frac{%
r}{\ell }=\sin \psi <\frac{\sqrt{3}}{2}$ we obtain for the spatial part of
de Sitter the line element%
\begin{equation}
ds_{3}^{2}=\ell ^{2}\left( d\psi ^{2}+\sin ^{2}\psi \left( d\theta ^{2}+\sin
^{2}\theta d\varphi ^{2}\right) \right)
\end{equation}%
which is compact $S^{3}$ metric with $-\sin ^{-1}\frac{\sqrt{3}}{2}<\psi <$ $%
\sin ^{-1}\frac{\sqrt{3}}{2}.$

ii) $\alpha =1,$ yields $r_{-}=R_{0},$ $r_{+}=3R_{0}$ in which the inner
horizon coincides with the shell. In this case we cast the spatial line
element into $S^{3}$ with $-\frac{\pi }{2}<\psi <$ $\frac{\pi }{2}.$ Note
that in this case our shell lies at the inner (Cauchy) horizon $r_{-}$ of RN
so that it becomes null which we shall not elaborate on. Instability of the
Cauchy horizon suggests \cite{Cauchy} that we must exclude this choice.

iii) $\alpha =\sqrt{3},$ gives $r_{-}=3R_{0}\left( 2-\sqrt{3}\right) $ and $%
r_{+}=3R_{0}\left( 2+\sqrt{3}\right) .$ For this case and in general for any 
$\alpha >1,$ we have $\frac{r}{\ell }>1$ so that the coordinate change $%
\frac{r}{\ell }>\cosh \tau $ and $t=\ell \rho $ yields%
\begin{equation}
\frac{ds^{2}}{\ell ^{2}}=-d\tau ^{2}+\sinh ^{2}\tau d\rho ^{2}+\cosh
^{2}\tau \left( d\theta ^{2}+\sin ^{2}\theta d\varphi ^{2}\right) .
\end{equation}%
This also gives a compact topology since $\tau $ is bounded from both below
and above as long as $\alpha $ stands for a finite number. Thus our result
is in conform with the Theorem of Borde \cite{Regularization}, which says
that regularity of a black hole demands that the topology changes. Let us
also add that in order to avoid the Cauchy horizon and be on the safe side
we must make the choice $\alpha >\sqrt{3}.$ Further, since the overcharging
case of the RN doesn't correspond to a black hole we ignore that discussion.
In \cite{9} they consider the case also with $Q^{2}>M^{2}.$

\section{The energy-momentum and Maxwell equations on the shell}

We shall explain in this section that although the energy-momentum $T_{\mu
}^{\nu }$ on the shell i.e., $S_{\mu }^{\nu }$ vanishes $\Delta T_{\mu
}^{\nu }=\lim_{r\rightarrow R_{0}^{+}}T_{\mu }^{\nu }-\lim_{r\rightarrow
R_{0}^{-}}T_{\mu }^{\nu }$ is not necessarily zero (A detailed discussion on
this issue is made by Bonnor and Vickers in \cite{BV}). This amounts to a
jump in $T_{\mu }^{\nu }$ across the shell. We have, for instance in the
present problem that%
\begin{equation}
\Delta T_{\theta }^{\theta }=\Delta T_{\varphi }^{\varphi }=\frac{6}{\ell
^{2}}
\end{equation}%
whereas $\Delta T_{t}^{t}=\Delta T_{r}^{r}=0.$ This is due to the fact that
the Israel junction conditions involve up to first order derivative of the
metric functions across the spherical shell while the angular energy
momentum tensor involves the second order derivatives of the metric
function. In other words, from the Israel junction condition we put
constraint on the metric functions and their first derivatives on the shell
while their second derivatives are free. Hence, while we set $\sigma
_{0}=p_{0}=0$ on the shell we only had to impose continuity condition on $%
f\left( r\right) $ and $f^{\prime }\left( r\right) $ across the shell but $%
f^{\prime \prime }\left( r\right) $ is free on the shell as we find%
\begin{equation}
\lim_{r\rightarrow R_{0}}\Delta f^{\prime \prime }\left( r\right) =\frac{12}{%
\ell ^{2}}.
\end{equation}%
For the discussion of Maxwell equation on the shell we make use of the
distributional potential 1-form given by 
\begin{equation}
A=Q\left( \frac{1}{r}-\frac{1}{R_{0}}\right) \Theta \left( r-R_{0}\right) dt
\end{equation}%
in which $\Theta \left( r-R_{0}\right) $ represents the Heaviside step
function and Q is the charge. Note that with this choice we assume that for $%
r>R_{0}$, up to a gauge transformation, we have the Coulomb potential $A_{0}=%
\frac{Q}{r}.$ The electromagnetic field 2-form becomes 
\begin{equation}
F=dA=\frac{Q}{r^{2}}\Theta \left( r-R_{0}\right) dt\wedge dr
\end{equation}%
where the notation is such that $d$ and $\wedge $ stand for the exterior
derivative and the wedge product, respectively. The dual 2-form of $F$ is
given accordingly by%
\begin{equation}
^{\ast }F=Q\sin \theta \Theta \left( r-R_{0}\right) d\theta \wedge d\varphi
\end{equation}%
so that the Maxwell equation on the shell takes the form 
\begin{equation}
d^{\ast }F=^{\ast }j.
\end{equation}%
Here $^{\ast }j$ is interpreted as the charge density $3-$form defined by 
\begin{equation}
^{\ast }j=Q\delta \left( r-R_{0}\right) \sin \theta dr\wedge d\theta \wedge
d\varphi
\end{equation}%
where $\delta \left( r-R_{0}\right) $ is the Dirac delta function coming
from the derivative of $\Theta \left( r-R_{0}\right) $ in the sense of
distributions. To show that Maxwell equation holds on the shell we check the
integral of $^{\ast }j$ which amounts to 
\begin{equation}
Q=\frac{1}{4\pi }\int \text{ }^{\ast }j
\end{equation}%
and is manifestly satisfied.

To complete this section let's add that having the Maxwell field non-zero on
one side of the shell must not lead to a conclusion that the energy momentum
on the shell can not be zero as the field is not extended to the other side
of the shell. We recall that the energy momentum on the shell is made by the
field from both sides of the shell. Hence, the presence of the cosmological
constant inside the shell guaranties that $p_{0}=\sigma _{0}=0$ provided (7)
and (8) are satisfied. As a matter of fact Eq.s (7) and (8) may be
interpreted as fine-tuning condition among parameters, mass, charge and the
cosmological constant. We see therefore that without the cosmological
constant inside the shell such a perfect match would not be possible,
justifying the choice of de Sitter as the inner spacetime.

\section{Stability of the model}

Once we adopt that the two spacetimes are glued on the timelike shell $%
F:=r-R=0$ we investigate next its stability. Here we assume a radial
perturbation of the shell which causes $R$ changing with respect to the
proper time $\tau .$ The standard calculation of the energy-momentum tensor
of the shell when $R=R\left( \tau \right) $ yields 
\begin{equation}
\sigma =-\frac{1}{4\pi G}\left( \frac{\sqrt{f_{2}\left( R\right) +\dot{R}^{2}%
}-\sqrt{f_{1}\left( R\right) +\dot{R}^{2}}}{R\left( \tau \right) }\right) ,
\end{equation}%
and%
\begin{multline}
p=\frac{1}{8\pi G}\left( \frac{2\ddot{R}\left( \tau \right) +f_{2}^{\prime
}\left( R\right) }{2\sqrt{f_{2}\left( R\right) +\dot{R}^{2}}}-\frac{2\ddot{R}%
\left( \tau \right) +f_{1}^{\prime }\left( R\right) }{2\sqrt{f_{1}\left(
R\right) +\dot{R}^{2}}}\right. + \\
\left. \frac{\sqrt{f_{2}\left( R\right) +\dot{R}^{2}}-\sqrt{f_{1}\left(
R\right) +\dot{R}^{2}}}{R\left( \tau \right) }\right) .
\end{multline}%
in which $f_{1}$ and $f_{2}$ are given in (5) and (6) and a dot represents $%
\frac{d}{d\tau }$. We note that the energy conservation equation imposes
that $\sigma $ and $p$ given in (19) and (20) satisfy%
\begin{equation}
\frac{d\sigma }{dR}+\frac{2}{R}\left( p+\sigma \right) =0.
\end{equation}%
An equation of state in the form of $p=p\left( \sigma \right) $ in this
equation manifests the exact form of $\sigma $ and $p$ after the
perturbation irrespective of the form of $f_{1}$ and $f_{2}.$ The latter
equation admits%
\begin{equation}
\int_{0}^{\sigma }\frac{d\sigma }{p\left( \sigma \right) +\sigma }=2\ln
\left( \frac{R_{0}}{R}\right)
\end{equation}%
which suggests that $p\left( \sigma \right) $ can not be an arbitrary
function as it must satisfy $p\left( 0\right) =0.$ Note that the integration
constant $R_{0}$ is identified as the unperturbed radius of the shell. For
instance a linear gas with EoS $p=\omega \sigma $ ($\omega =const.$) can not
be a physical choice. We recall that a massive shell was chosen in \cite{10}
which is different from our choice. An equation of state of the form 
\begin{equation}
p=-\sigma +\omega \sigma ^{\nu }
\end{equation}%
in which $0<\nu <1$ is a suitable candidate for the fluid presented on the
surface of the shell after the perturbation. This is a modified version of a
polytropic fluid \cite{Polytropic}. We note that the non-zero $p$ and $%
\sigma $ after the perturbation can be attributed to the energy given during
the perturbation. Obviously it is observed from (22) that $p=0$ when $\sigma
=0.$ Integrating (22) with (23) one finds%
\begin{equation}
\sigma \left( R\right) =\left( 2\omega \left( 1-\nu \right) \ln \left( \frac{%
R_{0}}{R}\right) \right) ^{\frac{1}{1-\nu }}.
\end{equation}%
Herein $\omega $ is a constant which can be adjusted but as $R$ gets values
on both sides of $R_{0}$ one has to set $\nu $ in such a way that the
right-hand side remains real. For instance $\nu =\frac{1}{2}$ leaves the
expression real while $\nu =\frac{3}{5}$ does not. The total energy on the
shell can be obtained as 
\begin{multline}
E=\int \sigma \left( R\right) \delta \left( r-R\right) \sqrt{-g}d^{4}x= \\
4\pi R^{2}\left( 2\omega \left( 1-\nu \right) \ln \left( \frac{R_{0}}{R}%
\right) \right) ^{\frac{1}{1-\nu }}.
\end{multline}%
In Fig. 2 we plot $E$ versus $R$ for $\omega =1,$ $\nu =\frac{1}{2}$ and for
the three different $R_{0}$ values used in Fig. 1. In accordance with Fig.
2, for some extension, more deviation from $R=R_{0}$ requires more energy
and physically this is an indication of stability. From Fig. 2 it is also
seen that the minimum of energy formed at $R=R_{0}$ is strongly stable from
right side. From the left, on the other hand, overcoming the energy barrier
causes the shell to collapse leaving behind a flat spacetime in accordance
with (7) and (8).

To justify the polytropic property, i.e., $PV^{n}=const.,$ (with $n=const.$)
of the equation of state (23) we choose a particular parameter, namely, $\nu
=\frac{1}{2}$ and make analysis in the vicinity of $R=R_{0}.$ With $\nu =%
\frac{1}{2}$ we have from (23)%
\begin{equation}
p=\omega ^{2}\ln \frac{R_{0}}{R}\left( 1-\ln \frac{R_{0}}{R}\right) .
\end{equation}%
Now we take $R=R_{0}+\epsilon ,$ where $\left\vert \epsilon \right\vert \ll
1 $ and upon expansion we obtain 
\begin{equation}
p\simeq -\frac{\epsilon \omega ^{2}}{R_{0}}.
\end{equation}%
Recalling that the volume, (in fact the area) $V\sim R_{0}^{2}$ for $S^{2}$
we have $PV^{1/2}\simeq -\epsilon \omega ^{2}=const.$ so that it corresponds
to $n=\frac{1}{2}$ law for the polytropic gas on the shell.

\section{Conclusion and discussion}

By applying the cut and paste technique via a thin-shell we regularize the
inner part of the RN spacetime which removes its central singularity. Simply
the patched regular de Sitter spacetime constitutes the inner part. This
amounts to the choice of the distributional metric function $f\left(
r\right) =\left( 1-\frac{r^{2}}{\ell ^{2}}\right) \Theta \left( R-r\right)
+\left( 1-\frac{2m}{r}+\frac{Q^{2}}{r^{2}}\right) \Theta \left( r-R\right) ,$
in which $R$ stands for the radius of the shell. At the static case $%
R=R_{0}, $ application of the Israel junction conditions yields no source on
the surface of the shell for a smooth match. This requires that the metric
and its first derivative are continuous on the shell which amounts to $M=%
\frac{2R_{0}^{3}}{\ell ^{2}}$ and $Q^{2}=\frac{3R_{0}^{4}}{\ell ^{2}}$ and
upon these identifications the second derivative $f^{\prime \prime }\left(
r\right) $ yields no Dirac delta function. This means that the shell is
source free i.e. $p_{0}=\sigma _{0}=0$ at the equilibrium condition and
these conditions emerge as a result of fine-tuning of parameters via (7) and
(8). However, upon radial perturbation we can have for both $R>R_{0}$ and $%
R<R_{0}$ a source of modified polytropic fluid. In this sense the shell may
be considered as a 'false vacuum state' for the environmental fluid
described by the equation of state (23) whose limit $R\rightarrow R_{0}$
agrees with such a vacuum. Relying on the curves of energy versus $R$ we
predict a restricted stability of the shell which makes the model feasible
to certain extend. It should also be added that the boundary shell must be
finely tuned to avoid the Cauchy horizon and its inherent instability.
Finally, we must add that RN singularity is a time-like one (for $M>Q$)
which may be considered weaker than the spacelike singularity of the
Schwarzschild black hole. Although our method has no immediate answer for
the removal of the Schwarzschild's singularity what we have shown in this
study is that in the case of RN it remarkably works through a change in
topology. This is due to the fact that the de-Sitter geometry must have only
a compact topology which confirms a theorem proved in \cite{Regularization}.
We add that recent methods of removing the singularity of black holes are
available in the literature which also are based on the topology change (see 
\cite{Acta} and the references cited therein).

\bigskip


\begin{thebibliography}{99}
\bibitem{Israel} G. Darmois (1927) M\'{e}morial de Sciences Math\'{e}%
matiques, Fascicule XXV, \textquotedblleft Les equations de la gravitation
einsteinienne\textquotedblright , Chapitre V.

W. Israel "Thin shells in general relativity". Il. Nuovo Cim. \textbf{66}, 1
(1966).

\bibitem{MV} M. Visser, Phys. Rev. D \textbf{39}, 3182 (1989).

\bibitem{MT} M. S. Morris and K. S. Thorne, Am. J. Phys. \textbf{56}, 395
(1988).

\bibitem{Maldacena} J. Maldacena and L. Susskind, Fortsch. Phys. \textbf{61}%
, 781 (2013).

\bibitem{ER} A. Einstein and N. Rosen, Phys. Rev. \textbf{48}, 73 (1935).

\bibitem{Vilenkin1} A. V. Vilenkin and P. I. Fomin, Nuovo Cimento Soc. Ital.
Fis., A \textbf{45}, 59 (1978).

\bibitem{Vilenkin2} A. V. Vilenkin and P. I. Fomin, ITP Report No.
ITP-74-78R, 1974.

\bibitem{Zaslavskii} O. B. Zaslavskii, Phys. Rev. D \textbf{70}, 104017
(2004).

\bibitem{9} J. P. S. Lemos and V. T. Zanchin, Phys. Rev. D \textbf{83},
124005 (2011).

\bibitem{10} N. Uchikata, S. Yoshida and T. Futamase, Phys. Rev. D \textbf{86%
}, 084025 (2012).

\bibitem{Frolov} V. P. Frolov, M. A. Markov and V. F. Mukhanov, Phys. Rev. D 
\textbf{41}, 383 (1990).

\bibitem{Frolov2} V. P. Frolov, Phys. Rev. D \textbf{94}, 104056 (2016).

\bibitem{RBH} J. Bardeen, Proceedings of GR 5 (Tbilisi, USSR, 1968).

\bibitem{Regularization} A. Borde, Phys. Rev. D \textbf{55}, 7615 (1997).

\bibitem{Wheeler} John A. Wheeler (1962) "Geometrodynamics" Acad. Press.

\bibitem{Cauchy} R. A. Matzner, N. Zamorano, and V. D. Sandberg, Phys. Rev.
D \textbf{19}, 2821 (1979).

\bibitem{BV} W. B. Bonnor and P. A. Vickers, General Rel. Gra. \textbf{13},
29 (1981).

\bibitem{Polytropic} M. Azam, S. A. Mardan, I. Noureen and M. A. Rehman,
Eur. Phys. J. C \textbf{76}, 315 (2016).

\bibitem{Acta} F. R. Klinkhamer and C. Rahmede, Phys. Rev. D \textbf{89},
084064 (2014);

F. R. Klinkhamer, Acta Phys. Pol. B \textbf{45}, 5 (2014);

F. R. Klinkhamer, Mod. Phys. Lett. A \textbf{29}, 1430018 (2014);

F. R. Klinkhamer, Mod. Phys. Lett. A \textbf{28}, 1350136 (2013).
\end{thebibliography}
\end{document}